\documentclass[letterpaper, 10 pt, conference]{ieeeconf}  

\IEEEoverridecommandlockouts                              

\usepackage{graphics} 
\usepackage{epsfig} 
\usepackage{mathptmx} 
\usepackage{times} 
\usepackage{amsmath} 
\usepackage{amssymb}  
\usepackage{cite}
\usepackage{amsmath,amssymb,amsfonts}
\usepackage{algorithmic}
\usepackage{graphicx}
\usepackage{textcomp}
\usepackage{xcolor}
\newcommand{\revI}[1]{\textcolor{black}{#1}}

\usepackage[ruled,vlined]{algorithm2e}
\usepackage{algorithmic}
\usepackage{tikz}
\usepackage{pgfplots}
\pgfplotsset{compat=newest} 
\pgfplotsset{plot coordinates/math parser=false} 
\usetikzlibrary{plotmarks,patterns,decorations.pathreplacing,backgrounds,calc,arrows,arrows.meta,spy,matrix}
\usepgfplotslibrary{patchplots,groupplots}
\usepackage{tikzscale}
\usepackage{cite}
\usepackage{hyperref}
\usepackage{makecell}

\newenvironment{IEEEkeywords}{\noindent\textbf{Keywords:}}{}

\usepackage[acronym]{glossaries}
\newacronym{DT}{DT}{Digital-Twin}
\newacronym{PL}{PL}{Power Loss}
\newacronym{pls}{PLs}{Power Losses}
\newacronym{plm}{PLM}{Power Loss Model}
\newacronym{TM}{TM}{Thermal Model}
\newacronym{fem}{FEM}{Finite Element Method}
\newacronym{ROM}{ROM}{Reduced Order Model}
\newacronym{LUT}{LUT}{Look Up Table}
\newacronym{BIBO}{BIBO}{Bounded Input Bounded Output}
\newacronym{fnn}{FNN}{Feed-Forward Neural Network}
\newacronym{RNN}{RNN}{Recurrent Neural Network}
\newacronym{mlp}{MLP}{Multi-Layer Perceptron}
\newacronym{ml}{ML}{machine learning}
\newacronym{poi}{PoI}{points of interest}
\newacronym{mse}{MSE}{mean squared error}
\newacronym{sgd}{SGD}{stochastic gradient descent}
\def\BibTeX{{\rm B\kern-.05em{\sc i\kern-.025em b}\kern-.08em
    T\kern-.1667em\lower.7ex\hbox{E}\kern-.125emX}}

\title{\LARGE \bf
Data-driven Power Loss Identification through \\
Physics-Based Thermal Model Backpropagation
}

\author{Mattia Scarpa$^{1,2}$ and Francesco Pase$^{2}$ and Ruggero Carli$^{1}$ and Mattia Bruschetta$^{1}$ and Francesco Toso$^{2}$
\thanks{$^{1}$Mattia Scarpa, Ruggero Carli, and Mattia Bruschetta are with the Faculty of Information Engineering,
        University of Padova, Padova, Italy
        {\tt\small \{scarpamatt, carlirug, bruschet\}@dei.unipd.it}}%
\thanks{$^{2}$Mattia Scarpa, Francesco Pase, and Francesco Toso are with Newtwen, Padova, Italy
        {\tt\small \{mattia.scarpa, francesco.pase, francesco.toso\}@newtwen.com}}%
}

\begin{document}

\maketitle
\thispagestyle{empty}
\pagestyle{empty}
\maketitle

\begin{abstract}
Digital twins for power electronics require accurate power losses whose direct measurements are often impractical or impossible in real-world applications. This paper presents a novel hybrid framework that combines physics-based thermal modeling with data-driven techniques to identify and correct power losses accurately using only temperature measurements. Our approach leverages a cascaded architecture where a neural network learns to correct the outputs of a nominal power loss model by backpropagating through a reduced-order thermal model. We explore two neural architectures, a bootstrapped feedforward network, and a recurrent neural network, demonstrating that the bootstrapped feedforward approach achieves superior performance while maintaining computational efficiency for real-time applications. Between the interconnection, we included normalization strategies and physics-guided training loss functions to preserve stability and ensure physical consistency. Experimental results show that our hybrid model reduces both temperature estimation errors (from 7.2±6.8°C to 0.3±0.3°C) and power loss prediction errors (from 5.4±6.6W to 0.2±0.3W) compared to traditional physics-based approaches, even in the presence of thermal model uncertainties. This methodology allows us to accurately estimate power losses without direct measurements, making it particularly helpful for real-time industrial applications where sensor placement is hindered by cost and physical limitations.
\end{abstract}
\vspace{1mm}

\begin{tikzpicture}[remember picture,overlay]
		\node[anchor=north,yshift=-12pt] at (current page.north) {\parbox{\dimexpr\textwidth-\fboxsep-\fboxrule\relax}{
				\centering\footnotesize This paper has been accepted for presentation at the 23rd IEEE European Control Conference \textcopyright 2025 IEEE. \\
				}};
	\end{tikzpicture}

\section{Introduction}
Thermal management and sensing play a critical role in many industrial applications that rely on power electronics. This need is found in several components, such as electric motors, inverters, and on-board chargers.
However, the design and implementation of solutions that require sensor placement in such components often face significant challenges, mainly due to the high cost and physical constraints that limit access in critical positions. These limitations often translate into fewer sensors being utilized, leading to a reduced system observability hindering the monitor, the control, and the maintenance of electrical power devices \cite{castillo_tapia_constrained_2023_SensorP}.

\gls{DT} has emerged as a crucial technology for advancing industrial applications, particularly in power electronics where thermal management plays a critical role \cite{tao_digital_2019_DT-SoA}. The integration of \gls{DT}s in thermal management applications has become increasingly important for monitoring, controlling, and maintaining electrical power devices \cite{chen_digital_2023_DT-PE}. However, implementing effective \gls{DT} solutions faces significant challenges, especially in real-time applications where accurate modeling of all system components is required \cite{chen_digital_2023_DT-PE, sanabria_challengeDT}. The main challenge in developing accurate digital twins for thermal management is the precise characterization of power losses, which refers to all forms of energy dissipation that manifest as heat in power electronics systems, mainly due to manufacturing and physical uncertainties \cite{torchio_multiphysics_2024, sanz-alcaine_estimation_2023}.
These limitations are particularly significant in power electronics applications, where accurate thermal management is crucial for device reliability, performance optimization, and control.

\subsection{Related Work}
The current approaches to power loss estimation either rely on physics-based or data-driven methods, or even a combination of the two. Traditional physics-based methods aim to estimate power losses through detailed analytical modeling and equivalent circuits \cite{kascak_method_2022, ma_013_2014}. While this approach provides good theoretical foundations, it is still limited due to the need for precise knowledge of the device parameters that may vary with the manufacturing process, making the framework less robust. Additionally, they might not address real-world variations and uncertainties. Recent advances in \gls{ml} have increased interest in data-driven solutions \cite{pl_est_1, pl_est_2}, demonstrating the potential of capturing non-linear relationships between input electro-thermal quantities and device losses \cite{nuzzo_adaptive_2024}.
Furthermore, training such approaches on empirical data enhances their generalization capabilities, enabling models to adapt to diverse operational conditions and devices from different manufacturers and manufacturing processes \cite{kirchgassner_hardcore_2024, da_silva_PL_ANN}.
Data-driven methodologies can be further combined with physics-based modeling, exploiting the device knowledge to enhance the overall model accuracy and robustness by learning missing system components that are not modeled using physics, or by identifying the parameters of a parameterized physical model \cite{li_data-driven_2022, solimene_hybrid_2024, kirchgassner_thermal_2023}. However, current data-driven solutions typically rely on having direct measurements of power losses during training or other feedback which are often unavailable in real applications. 
Despite remarkable advances in physics-based and data-driven approaches, power loss estimation in real-time remains an active research field, specifically when direct loss measurements are unavailable.

\subsection{Contribution}

In this work we propose a novel hybrid framework for cascaded systems in which the first component presents non-linear static behavior with limited model knowledge and no availability of output measurements. This subsystem is then connected to a dynamical system, accurately modeled through state-space representation, whose output measurements are accessible only for the training purpose but not to run in real-time. Considering this framework, our work leads to the following contributions:

\begin{itemize}
    \item A computationally efficient physics-based and data-driven modeling approach for thermal management.
    \item A methodology to learn and infer the physical discrepancies of the first subsystem using only indirect measurements from the second one, which guides the learning process through backpropagation. Moreover, a physics-informed loss has been designed to learn meaningful and physical predictions.
    \item A custom training strategy, which we call the \textit{bootstrap}  method, to correct the training process considering the dynamical aspects of the problem and the cascade architecture, which improves the overall accuracy of the model keeping the computational burden limited. 
    \item We then present extensive numerical results in support of our proposal by showing improved accuracy on the temperature estimations both in simulated and real environments, demonstrating the gain brought by the bootstrap approach, and by giving meaningful interpretations to the corrections made by the implemented neural networks.
\end{itemize}



\section{Model Description}
\label{sec:model_description}
We consider a typical thermal management system with two interconnected components: a \gls{plm} and a \gls{TM}. This cascade represents a common configuration in power electronics, particularly effective for physics-based Digital Twins\cite{torchio_multiphysics_2024}, which are demonstrating significant potentials for accurately modeling thermal dynamics with reliable temperature estimations.

\subsection{Thermal Model}
The Thermal Model (\gls{TM}) is formulated in a parameterized linear state space representation as follows:
\begin{equation}
\label{eq:TM}
    \Sigma_{TM}^{\vartheta}=\left(\mathbf{A}_\vartheta,\mathbf{B}_\vartheta,\mathbf{C}_\vartheta\right),
\end{equation}
where $\mathbf{A}_\vartheta\in\mathbb{R}^{n\times n}$, $\mathbf{B}_\vartheta\in\mathbb{R}^{n\times m}$, $\mathbf{C}_\vartheta\in\mathbb{R}^{p\times n}$ are the model matrices parameterized by $\vartheta$, which is the vector containing the physical parameters of the system. It is important to report that the state space model comes from high-fidelity \gls{fem} simulation and then, through a model reduction process, a \gls{ROM} is derived \cite{mor_benner_2021}. The reduction process preserves the system's most significant modes while reducing its complexity yet maintaining physical meaning and trading off accuracy with computational cost, which makes it suitable for real-time applications that must run on micro-controllers \cite{torchio_multiphysics_2024}. The \gls{TM} obtained with the reduction method is a parametric model whose parameters describe the device geometry and the properties of the specific material. The initial nominal values are denoted with $\tilde{\vartheta} \in \mathbb{R}^q$, which generates the state space model matrices.
These parameters require further calibration to ensure precise temperature predictions and high fidelity with the actual devices being modeled, bridging the gap from an as-designed to an as-manufactured model. Thus, the goal is to tune the nominal parameters $\tilde{\vartheta}$ aligning them as closely as possible with the real and unknown parameters vector $\vartheta^* \in \mathbb{R}^q$. 
Moreover, at the end of the calibration process, observers such as Kalman's filters can be adapted to improve accuracy based on the available temperatures in the real-time application.
\begin{figure}
    \centering
    \includegraphics[width=.8\linewidth]{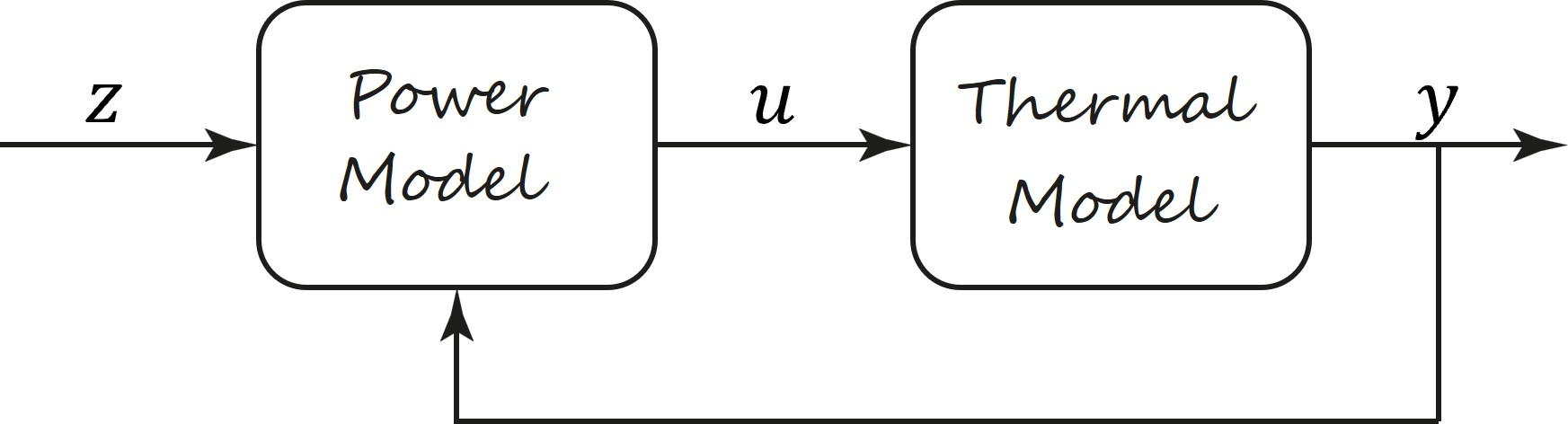}
    \caption{Power Loss Model and Thermal Model interconnected in cascade for temperature estimation.}
    \label{fig:cascade_pbdt}
\end{figure}

\subsection{Power Loss Model}
\gls{pls} are typically more challenging to be accurately modeled. Unlike dynamic \gls{TM}, they are usually approximated by some non-linear functions or \glspl{LUT}, mapping inputs such as currents, voltages and temperatures, to the meaningful losses of the device. In our configuration, the \gls{plm} is represented by a parameterized function

\begin{align}
\label{eq:PL}
    g_{PL}^{\varphi}:\mathbb{R}^d &\to \mathbb{R}^m \\
    g_{PL}^{\varphi}(\mathbf{z})&=\mathbf{u}
\end{align}
where $\varphi \in \mathbb{R}^h$ denotes the parameters vector containing information about resistance values, resistivity, and other physical coefficients, $d$ is the dimension of the input vector, while $m$ of the output one, which is also the input of the \gls{TM}. However, the estimation of the real parameters vector $\varphi^*$ brings several challenges, mainly due to:
\begin{itemize}
    \item \textbf{Non-Linearity}: the relationships between the inputs and the losses are often non-linear, making it difficult to capture them with simple analytical models.

    \item\textbf{Parameters Uncertainties}: The nominal value of the parameters vector $\tilde{\varphi}$ are inaccurate due to strong, and often inaccurate, assumptions and approximations on manufacturing tolerance, device materials properties and geometry. Consequently, the nominal model is unable to represent all non-linear physical phenomena, especially those coming from unmodeled dynamics and environmental coupling effects.

    \item \textbf{Measurement limitation}: Power loss measurements are typically not available, leading to relying only on indirect feedback.

    \item \textbf{Environmental dependence}: Power losses are also sensitive to environmental conditions, e.g. boundary conditions, which can vary significantly in real-world applications.
\end{itemize}
These factors contribute to higher uncertainty in the \gls{plm} model compared to the \gls{TM}, justifying our focus on the correction and improvement of the losses estimates.

\subsection{Model Interconnection}
The complete model, shown in \autoref{fig:cascade_pbdt}, can be represented as a cascade:
\begin{equation}
\label{eq:timePL}
    \mathbf{u}_k = g_{PL}^\varphi(\mathbf{z}_k),
\end{equation}
\begin{equation}
\label{eq:ss_dynamics}
    \biggl\{\begin{array}{l}
        \mathbf{x}_{k+1}=\mathbf{A}_\vartheta\mathbf{x}_k+\mathbf{B}_\vartheta\mathbf{u}_k \\
        \mathbf{y}_k = \mathbf{C}_\vartheta\mathbf{x}_k
    \end{array},
\end{equation}
where $\mathbf{z}_k \in \mathbb{R}^d$ represents the input vector, $\mathbf{u}_k \in \mathbb{R}^m$ the estimated power losses, and $\mathbf{y}_k \in \mathbb{R}^p$ the resulting temperatures in the \gls{poi}. This cascaded structure allows us to isolate each step of the temperature estimation process, potentially acting separately on the \gls{plm} and on the \gls{TM}.

\section{Methodology}
\label{sec:methodology}
In this section, we will explain in-depth how the challenge of estimating power losses has been improved resorting to a data-driven approach.

\subsection{Problem Formulation}
In this work, we aim to address the inaccuracies of the \gls{plm} parameters, which affect the input of the \gls{TM}, potentially resulting in erroneous temperature estimates. We start by defining the error in the \gls{plm} parametrization in \autoref{eq:PL} as follows:
\begin{equation}
\label{eq:PL_error}
    \tilde{\varphi} = \varphi^* + \epsilon_{\varphi} \implies g_{PL}^{\tilde{\varphi}}(\mathbf{z}_k)=g_{PL}^{\varphi^*}(\mathbf{z}_k)+\epsilon_{\mathbf{u}_k}
\end{equation}
As discussed in Section \ref{sec:model_description}, we assume that the thermal behavior can be accurately captured through state-space representations derived from \gls{fem} simulations, which is a standard method. This approach leverages detailed knowledge of device geometry and material properties, enabling precise physics characterization despite the complexities inherent in power loss estimation. We thus rely on the following assumptions: 
\begin{itemize}
    \item \textit{Model Accuracy}: The thermal \gls{ROM} presents high fidelity, i.e., to a given input $u_k$ corresponds a similar outputs, expressed as 
    \begin{equation}
        \tilde{\vartheta} \approx \vartheta^* \implies \revI{\Sigma_{TM}^{\tilde{\vartheta}}\left(\mathbf{u}_k\right)=\hat{\mathbf{y}}_k} \approx \mathbf{y}_k,
    \end{equation}
    where $ \Sigma_{TM}^{\tilde{\vartheta}}$ represents the thermal system as defined in \autoref{eq:TM} whose dynamics can be computed using \autoref{eq:ss_dynamics}.

    \item \textit{Stability Properties}: The thermal \gls{ROM} possesses both asymptotic stability and \gls{BIBO} stability.

    \item \textit{Non-Observability}: The thermal model does not ensure observability properties, due to the lack of feedback measurements in real applications.
\end{itemize}

\subsection{Data-Driven Power Losses Correction}
\label{sub:data-drive-ploss-correction}
\begin{figure}
    \centering
    \includegraphics[width=.8\linewidth]{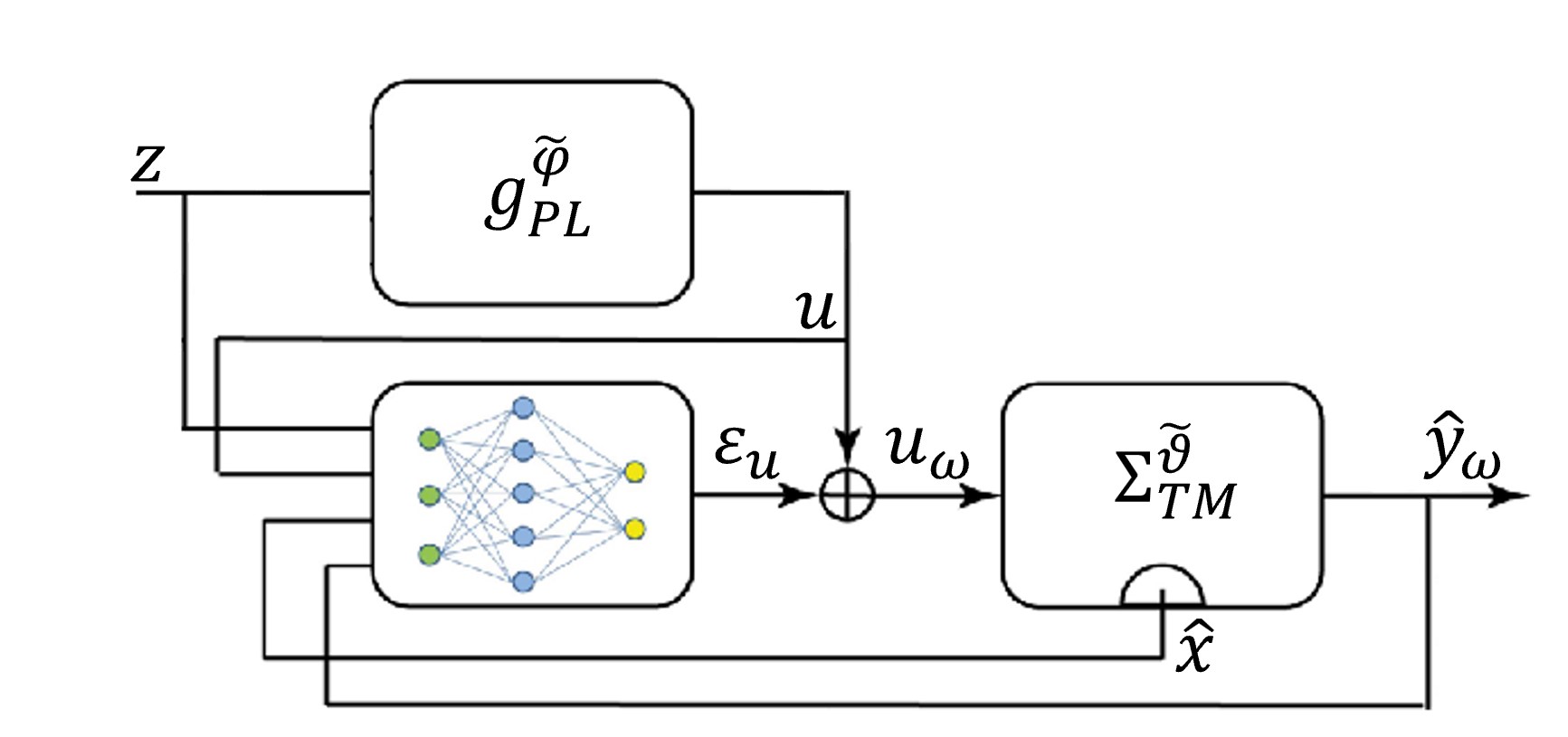}
    \caption{Data-Driven correction configuration that acts directly to correct the nominal estimated power losses.}
    \label{fig:nn_correction}
\end{figure}

To address the error of the power loss estimates, we propose a data-driven correction approach that acts directly on the \gls{PL} outputs by adopting a function $f_\omega\left(\cdot\right)$, which in our analysis is a neural network parameterized by ${\omega \in \mathbb{R}^{\Omega}}$, acting as a correction function. We thus leverage the accurate thermal model outputs to correct the losses estimate, constructing a hybrid physics-based data-driven architecture described in \autoref{fig:nn_correction}.
Consequently, our approach can be formalized as discovering $\omega^* \in \mathbb{R}^{\Omega}$ s.t. $\forall{k} \in \mathbb{N}$ it holds
\begin{align}
\begin{split}
\label{eq:datadriven_formalization}
    f_{\omega^*} \left(\mathbf{z}_k, \mathbf{u}_k,\hat{\mathbf{x}}_k,\hat{\mathbf{y}}_k, \dots\right) &= g_{PL}^{\varphi^*}(\mathbf{z}_k) - g_{PL}^{\tilde{\varphi}}(\mathbf{z}_k)
     = \epsilon_{\mathbf{u}_k}.
\end{split}
\end{align}
We remark that the only available measurements are the final temperatures $\mathbf{y}_k$, and not the real \gls{plm} outputs $\mathbf{u}_k^* = g_{PL}^{\varphi^*}(\mathbf{z}_k)$. Consequently, we need to find
\begin{equation}
\label{eq:base_loss}
    \omega^* = \text{argmin}_{\omega \in \mathbb{R}^{\Omega}} \: \mathcal{L} (\mathbf{y}, \hat{\mathbf{y}}_\omega)
\end{equation}
where the loss $\mathcal{L} (\mathbf{y}, \hat{\mathbf{y}}_\omega)$ is the \gls{mse} between the real temperatures $\mathbf{y}$ and the predicted ones $\hat{\mathbf{y}}_\omega$ obtained with the \gls{plm} parameterized by $\tilde{\varphi}$ and the correction function by $\omega$. The standard optimizers for deep learning models are within the family of \gls{sgd} algorithm like Adam \cite{Adam}, which is an iterative method that exploits the gradients of the cost function, i.e., $\mathcal{L}(\cdot)$ in our case, w.r.t. the parameters vector $\omega$ to minimize the problem. Gradients are usually estimated using the backpropagation algorithm \cite{dl_lecun_nature}, which back-propagates the gradients from the output of the model to each parameter in every layer. The idea of our proposal is to propagate the gradient information through the physics-based thermal model guiding the network updates to get physical meaning.

This work will explore and present two distinct neural network architectures to implement this correction strategy.

\subsubsection{Feedforward Neural Network Architecture}
The \gls{fnn}, known also as \gls{mlp}, has demonstrated remarkable effectiveness across diverse system identification problems and function approximation tasks. The Universal Approximation Theorem supports the theoretical foundations for their application in our context \cite{hornik_multilayer_1989}. For our power loss correction framework, we implement the correction function $f_\omega(\cdot)$ as an \gls{mlp} architecture. Each layer $l$ in the network performs the following transformation:
\begin{equation}
\label{eq:mlp}
    \mathbf{x}_{l+1}=\sigma\left(\mathbf{W}_l\mathbf{x}_l + \mathbf{b}_l\right),
\end{equation}
where $\mathbf{W}_l \in \mathbb{R}^{n_{l+1} \times n_l}$ represents the weight matrix, $\mathbf{b}_l \in \mathbb{R}^{n_{l+1}}$ denotes the bias vector, and $\sigma(\cdot)$ is a non-linear activation function \cite{dl_lecun_nature}. Despite their computational efficiency and demonstrated capabilities in various applications, straightforward application of \gls{fnn} architectures show significant limitations in capturing the intrinsic dynamics of cascaded thermal systems \cite{wang_forecast}, specifically in their generalization capabilities when learning in dynamical environments. 


To overcome these challenges, we elaborated custom training strategies focusing on the interconnection between the data-driven correction mechanism and the thermal model, resulting in two approaches that we included in our design: \gls{TM} normalization, and \gls{fnn} bootstrap during training.

\paragraph{Thermal Model Normalization}
As shown in \autoref{fig:nn_correction}, the \gls{fnn} includes in its inputs some \gls{TM} quantities, specifically the states and the temperatures estimates at the previous step and returns the right additive term to correct the \gls{TM} input. Due to the feedback interconnection, if even small errors are present in the corrected outputs, time integration potentially leads to a cascade effect that brings the network rapidly outside of its training domain, producing correction instabilities that reflect on the final temperature forecasting. To address this challenge we first exploit the \gls{BIBO} stability property of the \gls{TM} and set the activation function of the \gls{fnn} to be the hyperbolic tangent, i.e., $tanh$, whose output range is $[-1, 1]$ . This implies that the loss corrections, and consequently the temperature estimations, are always bounded.
However, despite the $tanh$ activation function advantages, it generally suffers from vanishing gradient problems. To overcome this, we normalized our thermal model and external inputs according to system operational conditions. From arbitrarily chosen $\mathbf{u}_{max}$ and $\mathbf{y}_{max}$, we define a transformation matrix $\mathbf{T}$ to obtain a new thermal model $\bar{\Sigma}_{TM}^{\vartheta}$ such that $\bar{\mathbf{u}}_{\omega} \rightarrow \bar{\Sigma}_{TM}^{\vartheta} \rightarrow \bar{\mathbf{y}}_{\omega}$
and $ \left|\bar{\mathbf{u}}_{\omega}\right| = \left|\frac{\mathbf{u}_{n}}{\mathbf{u}_{max}}\right| \&\leq 1$, 
$ \left|\bar{\mathbf{x}}_{\omega}\right| \&\leq 1$ and $ \left|\bar{\mathbf{y}}_{\omega}\right| \&\leq 1$.
The new thermal model always guarantees to have all its quantities bounded between $\pm1$, according to previously chosen maximum input/output values, mitigating the vanishing gradient issues maintained in every condition with reasonable and limited input.

\paragraph{Bootstrap Implementation}
As shown in \autoref{fig:cascade_pbdt}, the input-correcting neural network and the thermal model form a loop $f_{\omega} \rightarrow \textbf{\gls{TM}} \rightarrow f_{\omega}$. Consequently during training, for a given sequence of inputs $\{ \mathbf{z}_k\}_{k=1}^K$ and outputs $\{ \mathbf{y}_k\}_{k=1}^K$, the trajectory $\{ \mathbf{x}_k\}_{k=1}^K$ observed by the system changes as the neural network's parameters vector $\omega$ changes, thus the data used to train the network also change.
To cope with this during the training we periodically simulate the hybrid model and the new trajectories are collected for further training. The adopted strategy is better shown in \autoref{alg:bootstrap}. 
Specifically, at each epoch the function \texttt{predict()} will simulate the next instant model output starting from the data in the dataset $D$.
The loss function to be optimized is defined according to some criterion in \texttt{loss.compute()} and it will be used to compute the gradient exploiting backpropagation in \texttt{compute\_grads()}. These values will feed the optimizer that in \texttt{apply\_gradients()} will update the model parameters accordingly. Every $n_b$ epochs the model will simulate the new dynamics and in \texttt{add\_data()} will collect the new generated trajectories maintaining the train/validation/test division.
A fixed-memory buffer is also implemented to progressively allow forgetting the previous runs while injecting data with updated trajectories. As we will see the bootstrap improves significantly both the training and the forecast processes, reaching results comparable to neural networks with learnable dynamics, such as the \gls{RNN} (see Section \ref{subsec:rnn}), but using a simpler \gls{fnn} with reduced complexity.

\begin{algorithm}
\caption{Bootstrap Training Algorithm}
\label{alg:bootstrap}
\SetAlgoLined
\KwIn{Train dataset $D$, validation dataset $V$, number of epochs $E$, bootstrap update period $n_b$}
\vspace{2mm}
\hspace{1mm}\textbf{Initialize:} Hybrid Model $H\_model$, Optimizer $optimizer$, LossFunction $loss$\;
\vspace{2mm}
\For{$epoch = 1$ \KwTo $E$}{
    \If{$epoch \bmod n_b = 0$}{
        \hspace{1.8mm}$data \leftarrow H\_model.simulate()$\;
        $D.add\_data(data)$\;
    }
    \hspace{1.8mm}$train\_loss \leftarrow 0$\;
    $val\_loss \leftarrow 0$\;
    
    \ForEach{batch $b$ in $D$}{
        \hspace{1.8mm}$predictions \leftarrow H\_model.predict(b)$\;
        $batch\_loss \leftarrow loss.compute(predictions, b.labels)$\;
        $grads \leftarrow H\_model.compute\_grads(batch\_loss)$\;
        $optimizer.apply\_gradients(H\_model, grads)$\;
        $train\_loss \leftarrow train\_loss + batch\_loss$\;
    }
    \hspace{1.8mm}$val\_loss \leftarrow H\_model.evaluate(V)$\;
    $save(epoch, train\_loss, val\_loss)$\;
}

\textbf{return} Trained $H\_model$

\end{algorithm}

\subsubsection{Recurrent Neural Network Architecture}
\label{subsec:rnn}
Alternatively, to \gls{fnn} we also implemented a Recurrent Neural Network that, differently from the vanilla \gls{mlp}, introduces an additional internal state that allows modeling dynamics, a useful property for time series prediction.
Among the architectures that have been proposed in the past years, the most popular are Elman's \gls{RNN} \cite{elman_RNN}, Gated Recurrent Unit (GRU)\cite{gru2014}, and Long Short-Term Memory (LSTM)\cite{lstm1997}.
Recently, LSTM has been the most widely used architecture since it was designed to solve vanishing or exploding gradient problems. However, its additional complexity makes it unsuitable for maintaining computational efficiency to run the hybrid model in real-time on a constrained microprocessor. For this reason, our configuration will refer to the \gls{RNN} as Elman's Network derivation, and exploit the normalized \gls{TM} to mitigate gradient numerical problems. Additionally, the feature of managing temporal sequences allows different activation function choices, for instance, the \textit{Leaky-ReLU}, that already helps to address the vanishing gradient during training. The Elman's \gls{RNN} is formalized as
\begin{equation}
\label{eq:RNN_hidden}
    \mathbf{h}_k = \sigma(\mathbf{W}_{xh}\mathbf{x}_k + \mathbf{W}_{hh}\mathbf{h}_{k-1} + \mathbf{b}_h)
\end{equation}
\begin{equation}
\label{eq:RNNout}
    \mathbf{y}_k = \begin{cases}
        \mathbf{W}_{hy}\mathbf{h}_k + \mathbf{b}_y & \text{(single layer)} \\
        \mathcal{O}_{\omega'}(\mathbf{h}_k) & \text{(\gls{fnn})}
    \end{cases}
\end{equation}
where $\mathbf{h}_k$ represents the hidden state at step k, $\mathbf{x}_k$ denotes the input vector at step k, with $\mathbf{W}_{xh}$, $\mathbf{W}_{hh}$, and $\mathbf{W}_{hy}$ being respectively the input-to-hidden, hidden-to-hidden, and hidden-to-output weight matrices.
The terms $\mathbf{b}_h$ and $\mathbf{b}y$ are the bias vectors, while $\sigma(\cdot)$ represents the activation function. 
It is worth noticing that \autoref{eq:RNNout} has been slightly modified w.r.t. the original formulation as an additional \gls{fnn} from hidden-to-output layers can be adopted in case of strong non-linear relations.

\subsection{Loss Function Design}
To train our hybrid model we exploit the fact that all the operations are differentiable and therefore, we can use backpropagation between any possible input/output pair. 
While our primary objective was to minimize the power loss modeling error, direct loss measurements are typically unavailable in practical applications. Instead, we exploit thermal measurements from sensors carefully placed in the experimental/design phase, i.e., they are available during R\&D but not in production. Given these constraints, we formulate our base loss function on the available target temperature measurements (see Section \ref{sub:data-drive-ploss-correction}).
Although we expect the gradient information propagated through the thermal model to guide the parameter updates during training, the relative magnitudes of gradients corresponding to different model inputs play a crucial role.
If the gradient contributions are not balanced, this may lead to compensation effects in adjusting only some losses. This phenomenon can be particularly problematic as it may even result in physically inconsistent predictions, such as negative power losses, where the model artificially compensates for temperature discrepancies by violating fundamental physical constraints. Additionally, the non-observability assumption makes any reconstruction of the previous quantities only from the temperature output vector harder.  We chose to improve the training loss by adding two components to address these challenges. First, according to our framework, a nominal physical model of the \gls{plm} is available, providing a reference level for the power losses. We thus constrain the $L_2$-norm of our data-driven model's output to be within a factor $\zeta > 0$, representing the confidence of our nominal model. Consequently $\forall{k}$
\begin{align}
\label{eq:gap_assumption}
    \| g_{PL}^{\varphi^*}(\mathbf{z}_k) - g_{PL}^{\tilde{\varphi}}(\mathbf{z}_k) \|_2 \leq \zeta.
\end{align}
Therefore, we added regularization terms that penalize directly the correction, which corresponds to the network's output itself. Second, we added physical prior knowledge penalizing the losses $\mathbf{u}_{\omega}$ when negative. The final loss is thus:
\begin{equation}
\label{eq:improved_loss}
    \mathcal{L}_\omega=MSE(\mathbf{y}, \hat{\mathbf{y}}_\omega) + \alpha \ell (\mathbf{u}_\omega) + \beta\| \varepsilon_{u,\omega} \|_2
\end{equation}
where $\varepsilon_{u,\omega} = f_\omega\left(\cdot\right)$ indicates the output of the neural network, $\alpha > 0$ and $\beta>0$ are two regularization coefficients that must be carefully tuned during the optimization, and we introduced the penalty
\begin{align}
    \ell (\mathbf{u}_\omega) = \sum_i (\mathbf{u}_\omega)_i^2 \cdot \chi_{\{(\mathbf{u}_\omega)_i < 0\}}
\end{align}
in which $\chi_{(\mathbf{u}_\omega)_i < 0}$ is equal to $1$ when the i-th component is negative, and $0$ otherwise.
Temperature targets are collected from device trials characterized by a dense thermal sensor configuration extremely more comprehensive than the sensor configuration in production units.
While training dataset contains measurements recorded by many sensors placed on the device, we remind that only few physical sensors are placed during real-world operations, which can serve as input feedback for both the physics-based \gls{ROM} and the neural network.

\section{Numerical Results \& Conclusion}
\label{sec:results}

We designed an experimental framework aligned with our initial assumptions to validate the proposed architecture. Specifically, we maintain the constraint that only temperature measurements from the test bench are available for model training, and assume the larger source of uncertainties comes from the power loss parametrization and lack of information. 
We emphasize that temperature measurements used for the training comes from over-sensorized devices, which is a common procedure during the design phase, but they are not available in the real-time application.

\subsection{The Cascade Hybrid Model}

\begin{figure}
    \centering
    \includegraphics[width=.9\linewidth]{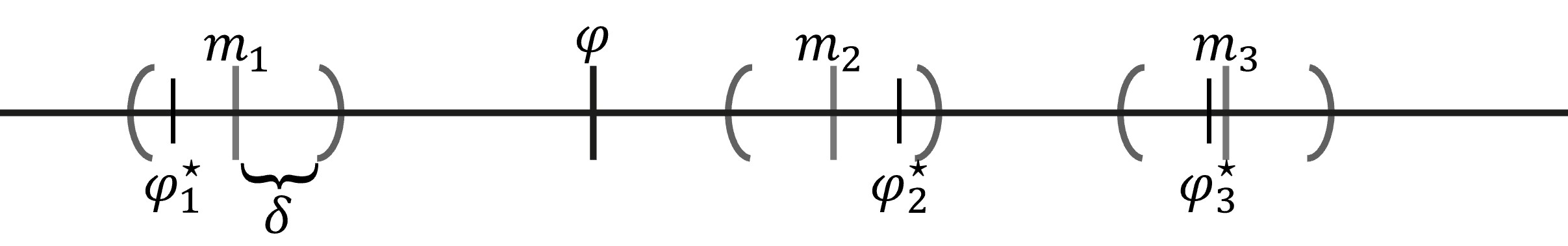}
    \caption{Distribution representation to select the simulated, unknown, real parameters from different distributions.}
    \label{fig:PL_error_params}
\end{figure}

To assess the quality of our loss corrections we implemented a simulation testbed in which we can directly observe and modify the real power losses by tweaking the parameters vector $\varphi$ of the power loss model. Specifically, we generated $n$ sets of parameters, as shown in \autoref{fig:PL_error_params}, simulating different operating conditions:
\begin{equation}
\label{eq:PL_params_generation}
    \varphi_i=\tilde{\varphi} + e_i, \hspace{3mm} e_i\sim\mathcal{U}\left(\left[m_i-\delta,m_i+\delta \right] \right) 
\end{equation}
where $m_i$ represents the average magnitude of the error for each experiment set $i$, which in our case was set to be $\pm50\%$ w.r.t. the original parameter, and $\mathcal{U}$ indicates the uniform distribution. Each of these error magnitudes in the \gls{plm} ideally symbolizes a different device, constructed from different manufacturing processes. The model is a non-linear mapping function that transforms electro-thermal quantities into $m=16$ distinct power losses. These estimated losses define the input vector for the \gls{ROM}, which predicts temperature behavior at $p=8$ \glspl{poi} within the electric device. For each new power loss parameter set, we simulate the model behavior under a series of current level inputs to generate the target temperatures $\mathbf{y}_i$ behavior. 
\renewcommand{\arraystretch}{1.3}  
\begin{table}[t]
    \centering
    \caption{Neural Network Architectures Parameters}
    \label{tab:nn_architectures}
    \begin{tabular}{|c|c|c|c|c|}
        \hline
        \makecell{Type} & \makecell{Input \\ Layer} & \makecell{Hidden \\ Layers/State} & \makecell{Output \\ Layer} & \makecell{Activation \\ Function} \\
        \hline
        \gls{fnn} & $13$ & $[15,25,15]$ & $16$ & $tanh$ \\
        \hline
        \gls{RNN} & $13$ & $25$ & $16$ & $leaky\_relu$ \\
        \hline
    \end{tabular}
\end{table}

To validate our proposed methodology we conducted experiments 
under different thermal model conditions. We specifically considered two different scenarios: The \textit{Accurate \gls{TM}}, in which the assumption $\tilde{\vartheta} = \vartheta^*$ holds, i.e., the thermal model is indeed accurate, and the \textit{Noisy \gls{TM}}, in which the vector components $\tilde{\vartheta}_i$ are generated by setting $\tilde{\vartheta}_i = \vartheta^*_i + \nu_i$ where $\nu_i$ is distributed according to a Gaussian  ${\nu_i \sim \mathcal{N}(0, \tau \cdot |\vartheta^*_i|)}$, with $\tau = 0.05$.

When designing the neural networks
we ensured an appropriate balance between model complexity and learning capacity, 
enhancing the physics-based \gls{plm}-\gls{TM} cascade by improving both temperature and power loss estimations while preserving physical consistency and real-time computational performance.
For this reason, both networks have a limited amount of layers and neurons, to respect the computational constraint. Their architectures and training parameters are detailed in \autoref{tab:nn_architectures}. The network input consists of input current, one temperature measurement from the available physical sensor, $p$ \gls{ROM}-estimated temperatures at the \gls{poi}s, and the first $3$ principal components obtained from PCA reduction of the \gls{ROM} states for a total of $nn_i=13$ inputs.
The output has size $m$ corresponding to the losses to be corrected.
The training consisted of a total of $n_{epoch}=6000$ epochs for all the networks, starting from a learning rate $lr=0.01$ under an exponential decay rate of $lr_{decay}=0.9999$. For the \gls{fnn} the model bootstrap was performed every $n_b=60$ epoch, and the sequence length for the \gls{RNN} training dataset was of $n_{seq}=50$ time steps.

\begin{figure}
    \centering
    \includegraphics[width=.9\linewidth]{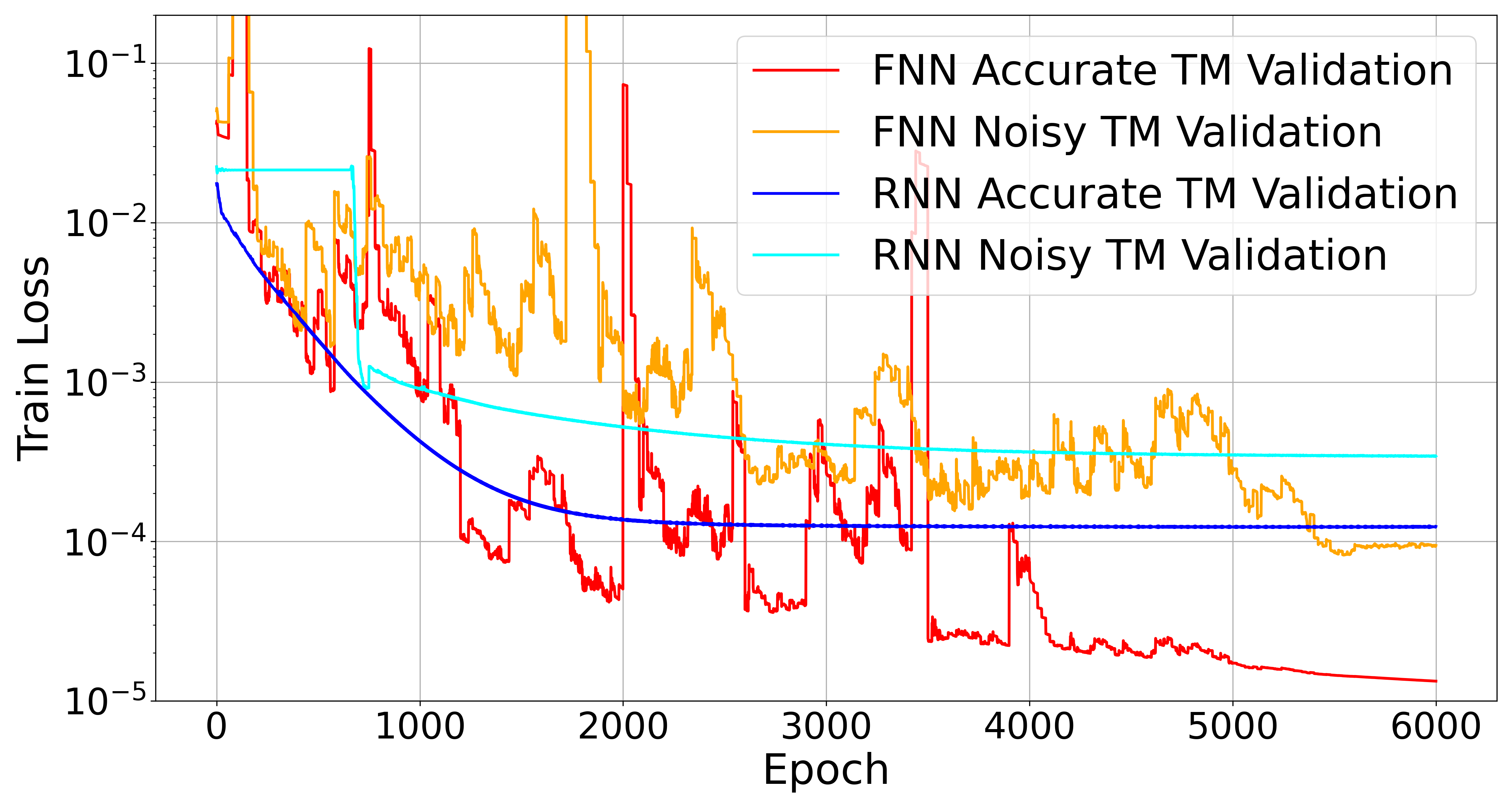}
    \caption{Hybrid model validation Losses for all the scenarios and configurations. The results show how the \gls{fnn} configuration with bootstrap can reach better performance with respect to a traditional \gls{RNN} in any scenario.}
    \label{fig:val_curve}
\end{figure}
\begin{figure}
    \centering
    \includegraphics[width=.9\linewidth]{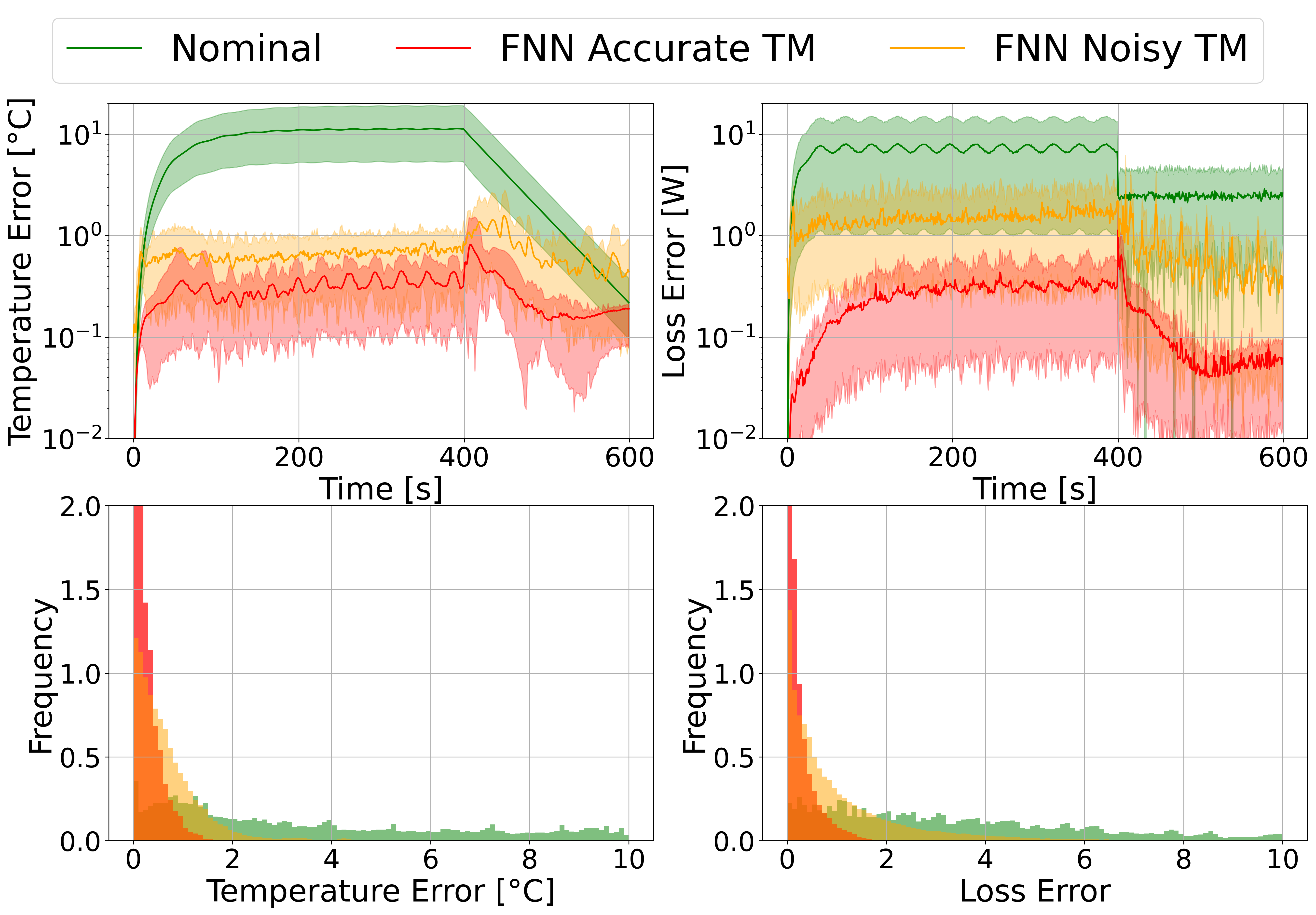}
    \caption{\gls{fnn} hybrid model comparison. Top: temperature and power loss estimation accuracy between physics-based and hybrid models. Bottom: absolute error distributions, showing accurate \gls{TM} outperforming noisy \gls{TM} by $58\%$ in temperature estimation and $82\%$ for power losses.}
    \label{fig:error_distrib}
\end{figure}

\subsection{Data Generation}
The dataset was generated by simulating a total of 10 models $(n_m=10)$, each with a different average $m_i$ (see Eq. (\ref{eq:PL_params_generation})), where $i \in[1,10]$. Each model received 6 current profile steps $(n_i=6)$ at different levels $L_i$, specifically
\begin{equation}
    \begin{array}{l}
         L_i\in[200, 400, 480, 600, 800] , \; \forall i\in[1,5] \\
         L_6 = 600 + sin(\frac{k}{10})
    \end{array}    
\end{equation}
The generated data were divided into three subsets: a training set, a validation set, and a test set. Specifically, $70\%$ of the data was allocated for training purposes, while $10\%$ was used to validate the models and select the best one based on the bias-variance tradeoff. The remaining $20\%$ of the data was derived from a set of unseen model parameters and was used to evaluate the accuracy of the models. We highlight that the $6$-th current profile was only used for testing purposes and none of its derived thermal dynamics was exploited for training the hybrid model.

\subsection{Results}
We now report the results achieved by our proposed Hybrid architecture and bootstrap training.
\textbf{Improved Accuracy.} The validation losses obtained in \autoref{fig:val_curve} demonstrate that the \gls{fnn} architecture with bootstrap implementation achieves significantly better performance compared to \gls{RNN} configurations in both scenarios. The bootstrap technique proved to play a crucial role in preventing error accumulation during forecasting, a common issue in cascaded interconnected systems. By periodically simulating and collecting new trajectories during training, bootstrapping enables the model to explore diverse operating conditions and system responses. This exploration mechanism enhances the model's ability to discover optimal solutions while preserving stability since it learns continuously from its predictions and adapts to different system behaviors.

Evaluating the error metrics we can observe significant improvements in both temperature and estimation accuracy of our hybrid model reported in \autoref{fig:error_distrib} with statistics in \autoref{tab:results}. 
For the temperature estimation, we can appreciate an improvement in the hybrid model accuracy of approximately $95\%$ w.r.t. the available accurate physics-based model. Also for the noisy scenario, we have obtained significant accuracy with an improvement of $\sim90\%$. These results are the direct consequence of our capability to accurately estimate the power losses, across all the test data resulting in a total improvement of about $\sim97\%$ with the accurate \gls{TM} and $\sim80\%$ for the noisy model.

\begin{figure}
    \centering
    \includegraphics[width=.75\linewidth]{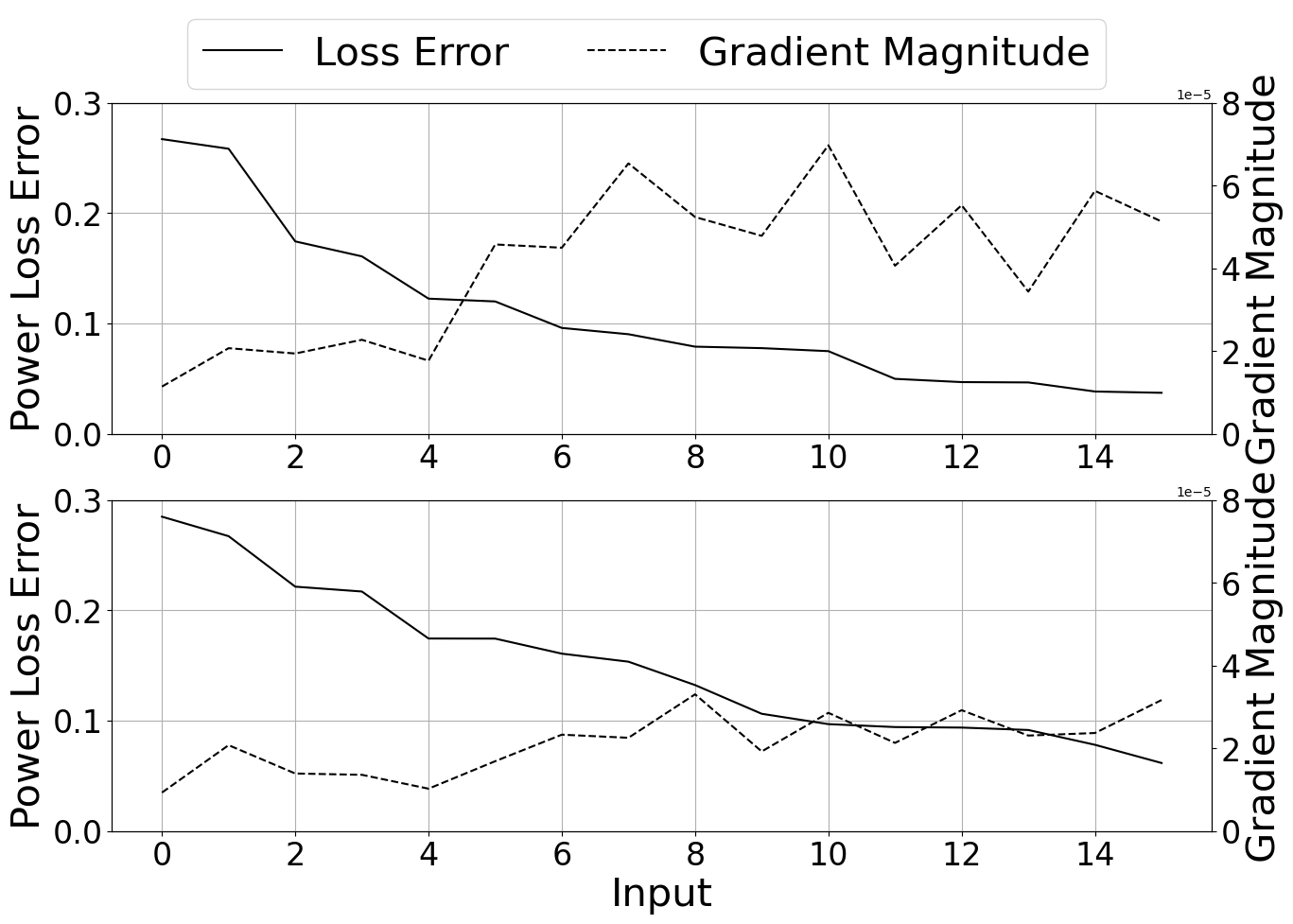}
    \caption{Estimation error of losses compared to the average gradient magnitude at each \gls{TM} (top accurate \gls{TM}, bot noisy \gls{TM}) input showing how for higher gradient during training corresponded lower losses estimation error.}
    \label{fig:error-grads_comparison}
\end{figure}

\textbf{Sensitivity Analysis}. Even when the achieved accuracies on the final outputs $\mathbf{y}$'s are very accurate (see \autoref{tab:results}), we observed that the errors on the power losses $\mathbf{u}^* - \mathbf{u}_\omega$ (which we can obtain only with our simulated toy model) were not always as small. To investigate this we exploit the gradients propagation through the \gls{TM} $\Sigma$ from its outputs to its inputs, which are computed during the training process and obtained \textit{for free} by automatic differentiation tools like PyTorch\cite{pytorch}. This way we performed a sensitivity analysis of $\Sigma$ \cite{sensitivity_power_2006}
investigating how much each input component $(\mathbf{u}_\omega)_i$ affects the outputs $\mathbf{y}$'s. \autoref{fig:error-grads_comparison} plots the relation between the gradient of the cost function on the temperatures w.r.t. the thermal inputs, i.e., $\frac{\partial \mathcal{L}_\omega(\mathbf{y}^*, \mathbf{y}_\omega)}{\partial \mathbf{u}_\omega}$, and the errors in the power losses reconstruction $\text{MSE}(\mathbf{u}^*, \mathbf{u}_\omega)$. It is possible to spot the correlation between the two: when the gradient magnitude is smaller, the reconstruction loss is higher, as further correcting that specific component has no impact on the final $\mathbf{y}$'s. These results corroborate our assumption of exploiting the \gls{TM} to provide physical information on the power losses corrections.

\renewcommand{\arraystretch}{1.3}  
\begin{table}[t]
    \centering
    \caption{Hybrid Model Estimation Performances}
    \label{tab:results}
    \begin{tabular}{|c|c|c|c}
        \hline
        \makecell{\textbf{Model}} & \makecell{\textbf{Temperature} \\ \textbf{Error}} & \makecell{\textbf{Loss} \\ \textbf{Error}} \\
        \hline
        \textit{Nominal Twin}      & $7.2\pm6.8$ & $5.4\pm6.6$  \\
        \hline
        \textit{Accurate \gls{TM}} & $0.3\pm0.3$ & $0.2\pm0.3$  \\
        \hline
        \textit{Noisy \gls{TM}}    & $0.7\pm0.6$ & $1.2\pm1.6$  \\
        \hline
    \end{tabular}
\end{table}

\begin{figure}
    \centering
    \includegraphics[width=.9\linewidth]{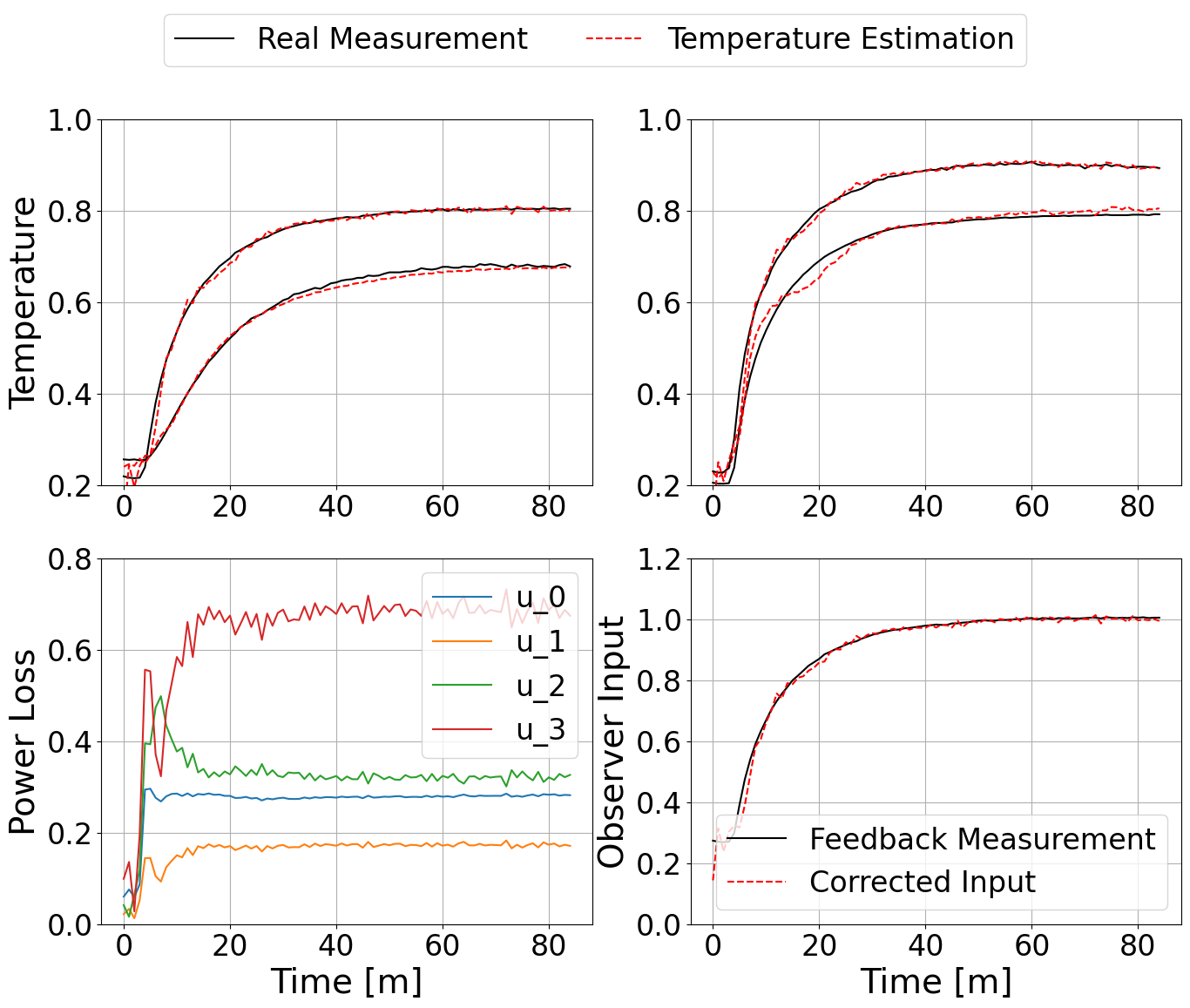}
    \caption{Real Scenario. In the top figures the real temperatures \revI{from two pairs of sensors} at \gls{poi}(solid line) and the hybrid model predictions (dotted line) are reported, showing how the loss corrections allow good estimation accuracy. The bottom left plot shows the corrected power loss remaining positive. In the bottom right plot, we observe the feedback measurement not modified by the network.}
    \label{fig:real_exp}
\end{figure}

\textbf{Real Scenario.} The final set of experiments regards the application of our hybrid scheme to a real-world problem. The task is to model the thermal behavior of an inlet charger for electrical vehicles exploiting the experimental data provided by our partner customer, for which the real parameters vector $\varphi^*$ of the power model does not exist (the parameterized is not known in general and was not assumed). Data have been normalized and anonymized for privacy reasons. The overall system can be modeled using the cascade scheme presented in \autoref{sec:model_description}, with the addition that the \gls{TM} in this case was equipped with a state observer exploiting the only sensor available in real-time acquiring the temperature in a specific device point. In \autoref{fig:real_exp} (top figures) the real measurements are plotted together with our model predictions, showing the high accuracy achieved by our proposed scheme. Furthermore, in the bottom-left plot in \autoref{fig:real_exp} the corrected power losses are reported to observe the obtained curves during inference, which are able to correct the biases of the nominal power loss made available by our partner. As mentioned, the thermal model gets as input also the measurement coming from one implemented sensor, which we also allowed to be corrected on purpose. The bottom-right plot in \autoref{fig:real_exp} shows that indeed the thermal input related to the observer was not modified by our implemented function $f_\omega$, as the two curves overlap, demonstrating the meaningful interpretation we can give to the correction mechanism, as that input contain highly-reliable information for the thermal model. 

\subsection{Conclusion}

Our experimental results show the effectiveness of the physics-based data-driven hybrid configuration improving both power loss and thermal dynamics estimation accuracy without requiring any direct loss measurements. The \gls{fnn} architecture implemented and trained with the proposed bootstrap technique seems to be the most effective configuration providing better generalization properties and better performances across different operating conditions and thermal uncertainties, maintaining also reduced complexity allowing the full hybrid model to run on embedded devices for real-time applications.
These results have significant implications for industrial applications where measurements, such as power losses, are often unavailable. 

Future works consists in the extension of this approach with different dynamic network architectures, and in the analysis on the stability between the interconnected subsystems.

\bibliographystyle{IEEEtran}
\bibliography{biblio}

\begin{thebibliography}{10}
\providecommand{\url}[1]{#1}
\csname url@samestyle\endcsname
\providecommand{\newblock}{\relax}
\providecommand{\bibinfo}[2]{#2}
\providecommand{\BIBentrySTDinterwordspacing}{\spaceskip=0pt\relax}
\providecommand{\BIBentryALTinterwordstretchfactor}{4}
\providecommand{\BIBentryALTinterwordspacing}{\spaceskip=\fontdimen2\font plus
\BIBentryALTinterwordstretchfactor\fontdimen3\font minus \fontdimen4\font\relax}
\providecommand{\BIBforeignlanguage}[2]{{%
\expandafter\ifx\csname l@#1\endcsname\relax
\typeout{** WARNING: IEEEtran.bst: No hyphenation pattern has been}%
\typeout{** loaded for the language `#1'. Using the pattern for}%
\typeout{** the default language instead.}%
\else
\language=\csname l@#1\endcsname
\fi
#2}}
\providecommand{\BIBdecl}{\relax}
\BIBdecl

\bibitem{castillo_tapia_constrained_2023_SensorP}
A.~Castillo~Tapia and A.~Román~Messina, ``Constrained sensor placement and state reconstruction in power systems from partial system observations,'' \emph{International Journal of Electrical Power \& Energy Systems}, 2023.

\bibitem{tao_digital_2019_DT-SoA}
F.~Tao, H.~Zhang, A.~Liu, and A.~Y.~C. Nee, ``Digital {Twin} in {Industry}: {State}-of-the-{Art},'' \emph{IEEE Transactions on Industrial Informatics}, 2019.

\bibitem{chen_digital_2023_DT-PE}
H.~Chen, Z.~Zhang, P.~Karamanakos, and J.~Rodriguez, ``Digital {Twin} {Techniques} for {Power} {Electronics}-{Based} {Energy} {Conversion} {Systems}: {A} {Survey} of {Concepts}, {Application} {Scenarios}, {Future} {Challenges}, and {Trends},'' \emph{IEEE Industrial Electronics Magazine}, 2023.

\bibitem{sanabria_challengeDT}
D.~M. Botín-Sanabria, A.-S. Mihaita, R.~E. Peimbert-García, M.~A. Ramírez-Moreno, R.~A. Ramírez-Mendoza, and J.~d.~J. Lozoya-Santos, ``Digital twin technology challenges and applications: A comprehensive review,'' \emph{Remote Sensing}, 2022.

\bibitem{torchio_multiphysics_2024}
\BIBentryALTinterwordspacing
R.~Torchio, F.~Toso, F.~Conte, J.~Ferretti, L.~Fazzini, N.~Matteazzi, P.~Bolognesi, and L.~Papini, ``Multiphysics {Thermal} {Digital} {Twin} of a {High} {Power} {Density} {Motor} for {Automotive} {Applications},'' in \emph{2024 International Conference on Electrical Machines (ICEM)}, 2024. [Online]. Available: \url{https://ieeexplore.ieee.org/document/10700506}
\BIBentrySTDinterwordspacing

\bibitem{sanz-alcaine_estimation_2023}
J.~M. Sanz-Alcaine, E.~Sebastian, F.~J. Perez-Cebolla, A.~Arruti, C.~Bernal-Ruiz, and I.~Aizpuru, ``Estimation of {Semiconductor} {Power} {Losses} {Through} {Automatic} {Thermal} {Modeling},'' 2023.

\bibitem{kascak_method_2022}
S.~Kascak and P.~Resutik, ``Method for estimation of power losses and thermal distribution in power converters,'' \emph{Electrical Engineering}, 2022.

\bibitem{ma_013_2014}
M.~Ke, A.~S. Bahman, S.~Beczkowski, and F.~Blaabjerg, ``{Loss} and {Thermal} {Model} of {Power} {Semiconductors} {Including} {Device} {Rating} {Information},'' in \emph{2014 International Power Electronics Conference (IPEC)}, 2014.

\bibitem{pl_est_1}
S.~Kalker, D.~Meier, C.~van~der Broeck, and R.~De~Doncker, ``Self-calibrating loss models for real-time monitoring of power modules based on artificial neural networks,'' \emph{IEEE Energy Conversion Congress and Expositio}, 2022.

\bibitem{pl_est_2}
C.~van~der Broeck, T.~Polom, and R.~De~Doncker, ``Degradation diagnosis of power modules based on thermal phase response sensing and artificial neural networks,'' \emph{IEEE Transaction on Industry Application}, 2024.

\bibitem{nuzzo_adaptive_2024}
G.~D. Nuzzo, A.~P. Pai, and Y.~Su, ``Adaptive {Artificial} {Neural} {Networks} for {Power} {Loss} {Prediction} in {SiC} {MOSFETs},'' in \emph{IEEE 10th Electronics System-Integration Technology Conference (ESTC)}, 2024.

\bibitem{kirchgassner_hardcore_2024}
W.~Kirchgässner, N.~Förster, T.~Piepenbrock, O.~Schweins, and O.~Wallscheid, ``{HARDCORE}: {H}-field and power loss estimation for arbitrary waveforms with residual, dilated convolutional neural networks in ferrite cores,'' 2024.

\bibitem{da_silva_PL_ANN}
G.~G. da~Silva, A.~de~Queiroz, E.~Garbelini, W.~P.~L. dos Santos, C.~R. Minussi, and A.~Bonini~Neto, ``Estimation of total real and reactive power losses in electrical power systems via artificial neural network,'' 2024.

\bibitem{li_data-driven_2022}
W.~Li, I.~Demir, D.~Cao, D.~Jöst, F.~Ringbeck, M.~Junker, and D.~U. Sauer, ``Data-driven systematic parameter identification of an electrochemical model for lithium-ion batteries with artificial intelligence,'' \emph{Energy Storage Materials}, 2022.

\bibitem{solimene_hybrid_2024}
L.~Solimene, C.~S. Ragusa, A.~Giuffrida, N.~Lombardo, F.~Marmello, S.~Morra, and M.~Pasquale, ``A {Hybrid} {Data}-{Driven} {Approach} in {Magnetic} {Core} {Loss} {Modeling} for {Power} {Electronics} {Applications},'' in \emph{{IEEE} 22nd {Mediterranean} {Electrotechnical} {Conference} ({MELECON})}, 2024.

\bibitem{kirchgassner_thermal_2023}
W.~Kirchgässner, O.~Wallscheid, and J.~Böcker, ``Thermal neural networks: {Lumped}-parameter thermal modeling with state-space machine learning,'' \emph{Engineering Applications of Artificial Intelligence}, 2023.

\bibitem{mor_benner_2021}
\BIBentryALTinterwordspacing
P.~Benner, S.~Grivet-Talocia, A.~Quarteroni, G.~Rozza, W.~Schilders, and L.~Silveira, \emph{Model Order Reduction: Snapshot-Based Methods and Algorithms}.\hskip 1em plus 0.5em minus 0.4em\relax Berlin, Boston: De Gruyter, 2021. [Online]. Available: \url{https://doi.org/10.1515/9783110671490}
\BIBentrySTDinterwordspacing

\bibitem{Adam}
D.~P. Kingma and J.~L. Ba, ``Adam: A method for stochastic optimization,'' \emph{International Conference on Learning Representation}, 2015.

\bibitem{dl_lecun_nature}
Y.~LeCun, Y.~Bengio, and G.~Hinton, ``Deep learning,'' \emph{Nature}, 2015.

\bibitem{hornik_multilayer_1989}
K.~Hornik, M.~Stinchcombe, and H.~White, ``Multilayer feedforward networks are universal approximators,'' \emph{Neural Networks}, 1989.

\bibitem{wang_forecast}
R.~Wang, D.~Maddix, C.~Faloutsos, Y.~Wang, and R.~Yu, ``{Learning} {Dynamical} {Systems} {Requires} {Rethinking} {Generalization},'' in \emph{1st NeurIPS workshop on Interpretable Inductive Biases and Physically Structured Learning}, 2020.

\bibitem{elman_RNN}
J.~L. Elman, ``Finding structure in time,'' \emph{Cognitive Science}, 1990.

\bibitem{gru2014}
K.~Cho, B.~van Merrienboer, C.~Gulcehre, F.~Bougares, H.~Schwenk, and Y.~Bengio, ``Learning phrase representations using rnn encoder-decoder for statistical machine translation,'' \emph{Conference on Empirical Methods in Natural Language Processing (EMNLP 2014)}, 2014.

\bibitem{lstm1997}
S.~Hochreiter and J.~Schmidhuber, ``Long short-term memory,'' \emph{Neural Computation}, 1997.

\bibitem{pytorch}
A.~Paszke, S.~Gross, S.~Chintala, G.~Chanan, E.~Yang, Z.~De~Vito, Z.~Lin, A.~Desmaison, L.~Antiga, and A.~Lerer, ``Automatic differentiation in pytorch,'' \emph{31st Conference on Neural Information Processing Systems (NIPS 2017), Long Beach, CA, USA}, 2017.

\bibitem{sensitivity_power_2006}
I.~Hiskens and J.~Alseddiqui, ``Sensitivity, approximation, and uncertainty in power system dynamic simulation,'' \emph{IEEE Transactions on Power Systems}, 2006.

\end{thebibliography}

\end{document}